\documentstyle[12pt,aasms4]{article}

\newcommand{\rxte}{{\it RXTE\ }}
\newcommand{\batse}{{\it BATSE\ }}
\newcommand{\cgro}{{\it CGRO\ }}
\newcommand{\sgr}{{SGR~$1900+14$\ }}
\newcommand{\asca}{{\it ASCA\ }}

\begin{document}
\title{Discovery of a magnetar associated with the Soft Gamma
Repeater SGR~$1900+14$}

\author{C. Kouveliotou\altaffilmark{1,2}, T.
Strohmayer\altaffilmark{3}, K. Hurley\altaffilmark{4}, J. van
Paradijs\altaffilmark{5,6}, M.H. Finger\altaffilmark{1,2}, S.
Dieters\altaffilmark{5}, P. Woods\altaffilmark{5}, C.
Thompson\altaffilmark{7}, R.C. Duncan\altaffilmark{8}}
\authoremail{chryssa.kouveliotou@msfc.nasa.gov}

\altaffiltext{1} {Universities Space Research Association}
\altaffiltext{2} {NASA/Marshall Space Flight Center, ES-84, Huntsville,
AL 35812, USA}
\altaffiltext{3} {NASA/Goddard Space Flight Center, Greenbelt, MD
20771, USA}
\altaffiltext{4} {University of California, Berkeley, Space Sciences
Laboratory, Berkeley, CA 94720-7450, USA}
\altaffiltext{5} {Dept. of Physics, University of Alabama in
Huntsville, Huntsville, AL 35899, USA}
\altaffiltext{6} {Astronomical Institute ``Anton Pannekoek'', University of 
Amsterdam, The Netherlands}
\altaffiltext{7} {Department of Physics \& Astronomy, University of North
Carolina, Chapel Hill, NC 27599, USA} 
\altaffiltext{8} {Department of Astronomy, University of Texas, RLM
15.308, Austin, TX 78712,USA} 

\begin{abstract}

The soft-gamma repeater \sgr became active again on June 1998 after a
long period of quiescence; it remained at a low state of activity
until August 1998, when it emitted a series of extraordinarily intense
outbursts. We have observed the source with \rxte twice, during the
onset of each active episode. We confirm the pulsations at the 5.16 s
period reported earlier (\cite{hurley7001,khibid}) from \sgr. Here we
report the detection of a secular spindown of the pulse period at an
average rate of $1.1\times10^{-10}$ s/s. In view of the strong
similarities between SGRs, we attribute the spindown of \sgr to
magnetic dipole radiation, possibly accelerated by a quiescent flux,
as in the case of SGR~$1806-20$ (\cite{cknature}). This allows an
estimate of the pulsar dipolar magnetic field, which is $2-8 \times
10^{14}$ G. Our results confirm that SGRs are magnetars.

\end{abstract}


\section{Introduction}

Soft-gamma repeater sources (SGRs) were recognized as a new class of
objects in the late 80's (\cite{atteia87,laros87,ck87}). To date, only
four SGRs are confirmed; a fifth SGR source may have been detected
twice (\cite{hurley97,ck97,smith97}) but its existence is not yet
firmly established. What distinguishes these transient soft
$\gamma-$ray bursters from the `classic' $\gamma-$ray burst sources
is their recurrence and the much softer spectra of their bursts. SGRs
undergo random intervals of intense activity, during which
they emit bunches of up to hundreds of very short, low-energy bursts.
The bursts vary in duration and temporal structure from simple, single
pulses lasting less than 10 ms (\cite{ck95}), to long, highly complex
events that comprise over 40 very short subpulses each lasting tens of
ms (\cite{ck6929,khibid2}). The SGR burst arrival times seem to be distributed
lognormally, similar to the distribution of earthquakes
(\cite{epstein,hurley}).

Three SGRs have firm associations, based on positional coincidences,
with supernova remnants (SNRs). The position of SGR~$1900+14$, has
only recently been determined with sufficient accuracy
(\cite{khibid2}) to establish that it lies close to the SNR
G~$42.8+0.6$, whose association with \sgr was suggested earlier
(\cite{ck94,hurley94,vasisht94}). The SGR/SNR associations indicate
that SGRs are neutron stars, a conjecture strongly supported by the
detection of a three-minute train of 8 s pulsations following the
famous March 5, 1979 event from SGR~$0526-66$ (\cite{mazets}). 

Searches for pulsed emission in the SGR persistent flux were finally
successful when observations of SGR~$1806-20$ with the Rossi X-ray
Timing Explorer (\rxte) in November 1996 revealed a 7.47 s X-ray
pulsar (\cite{cknature}). \rxte observations of the newly discovered
SGR~$1621-47$ (\cite{ck6944}) indicated weak evidence of 6.7 s
pulsations from the source (\cite{dieters}). Finally, in recent
observations (April 1998) of \sgr Hurley et al. (1998e) detected
pulsed emission with 5.16 s period.  The pulsations of SGR~$1806-20$
show a secular spindown. As argued by Kouveliotou et al. (1998) this
spindown is due to magnetic dipole radiation; the corresponding
dipolar component of the magnetic field of SGR~$1806-20$ exceeds
$10^{14}$ G, thus establishing SGRs as `magnetars'
(\cite{duncan92,thompson95}). 

We report the results of observations with \rxte obtained soon after
the reactivation of the source was detected with the Burst And
Transient Source Experiment (\batse) on \cgro and Ulysses in May 1998
(\cite{hurley6929,ck6929}). Additional observations were obtained after the
recent extraordinary source activity of August 1998 (\cite{ron7002}),
when it emitted several events, one similar to the March 5, 1979 event
(\cite{cline7002,hurley7004,sax98}).  During these observations we
detected the 5.16 s pulse period (\cite{hurley7001,khibid}), and found that it
shows a secular spin-down at a rate similar to that observed for
SGR~$1806-20$ (\cite{ck7001}). The great similarities between SGRs
indicate that the causes of the spin-down of the pulsations in \sgr and
SGR~$1806-20$ are the same, i.e., \sgr is also a magnetar; its dipolar
magnetic field strength is $2-8\times10^{14}$ G.

\section{\rxte Observations}

\subsection{The May 1998 Observation}

We observed \sgr between May 31 and June 9, 1998 for a total of 41.7
ks on source\footnote{The start and end times of the individual
observations can be found at http://heasarc.gsfc.nasa.gov.}. After
excluding all bursts from the time series, the data were
energy-selected (2-20 keV), barycentered and binned at 0.25 s
resolution. We calculated a fast-Fourier power spectrum of the
resulting light curve, searching between 0-2.0 Hz for the 5.158975(7)
s period reported for \sgr from the \asca data (\cite{khibid}). Figure
1a shows the Power Density Spectrum (PDS) of the FFT; we can clearly
see the fundamental period at 5.159142(3) s (0.193831 Hz)
($JD~2450970.5$) and its first, second and third harmonics. The chance
probability of detecting this signal at the \sgr period is less than
$3\times10^{-30}$. This number is derived after we averaged the power
of the fundamental and the three harmonics between $0-1.0$ Hz, taking
into account the number of trials. We do not find significant power at
any other frequency; in particular, we do not detect signal at the
89.17 s (0.011 Hz) period from the X-ray pulsar XTE~J$1906+09$
reported earlier (\cite{marsden}) in the error box of \sgr. 

Figure 2a shows the epoch-folded PCA pulse profile; it is very similar
to the profile from the \asca data obtained roughly a month earlier
(\cite{khibid}), although the energy ranges are not identical (the
\asca range is 2-10 keV).  The two periods differ significantly; we
estimate the period derivative between the \asca and \rxte
observations to be $\dot{P} = 5.44(24)\times10^{-11}$ s/s. Using a pulse
folding analysis on the PCA data alone, in which the period is assumed
to be given by $P = P_0 + \dot{P} t$, we find that during the 10 days
of the PCA observations a significant change of the period occured.
The best fit is $\dot{P}=(1.10\pm0.17)\times10^{-10}$ s/s. We discuss
the difference between the two $\dot{P}$ values in Section 3.

\subsection{The August 1998 Observation}

\sgr started triggering \batse again on August 28, 1998. Prior to this
first trigger, the \rxte/All Sky Monitor had detected a rising
transient source from the same direction as \sgr (\cite{ron7002}). In
addition, the Ulysses and {\it KONUS-WIND\ } spacecraft recorded an
extremely intense event on 27 August, 1998, which was very similar to
the March 5, 1979 event (\cite{cline7002,hurley7004}); during this
event the source was occulted by the Earth for \batse (\cite{ck7003}).
A new series of ToO observations was initiated with \rxte\footnote{A
listing of the times of these observations can be found in
http://heasarc.gsfc.nasa.gov.}. We have analyzed 2.5 ks of on source
data obtained on August 28, 1998. From a pulse timing analysis we find
the pulsar period to be $P=5.160199(2)$ s (on $JD~2451056.5$); the
$\dot{P}$ during the August 1998 observation is
$1.14(23)\times10^{-10}$ s/s. Figure 1b shows the PDS of the FFT; the
fundamental and the first harmonic dominate this spectrum. The average
$\dot{P}$ between the first and second \rxte observations is
$1.406(5)\times10^{-10}$ s/s, 2.6 times the rate between the \asca and
the first \rxte observation. 

The folded pulse profiles obtained during 1998 May and August are
clearly different (see Fig. 2b), although they cover the same energy
range; the August 1998 profile is also different from that obtained
with \asca. This is reminiscent of SGR~$1806-20$ (\cite{cknature}),
for which we observed a spiky pulse profile with substructure when the
source was `inactive' (\asca observation), whereas a smooth, almost
sinusoidal profile appears {\it after} the source had become very
active (\rxte observation). The variable pulse profile of \sgr could
be ascribed to residual heat output in the active region following the
August 27 event.  

The evolution of $\dot{P}$ is shown in Figure 3, where we plot the
rate of spindown calculated between each consecutive observation as
well as within each \rxte observation. Note that the spindown rate
remains practically constant after June 1998.

\subsection{Search for Orbital Period}

In the August 1998 \rxte observations we found systematic departures
of pulse phases from the best fit ephemeris. These departures of $\sim
0.05$ cycles typically persisted for a few thousand seconds. Since
such signatures can potentially be due to binary orbital motion, we
searched for periodic variations in the pulse phases. 

We analysed 53 ks of {\it PCA\ } standard1 data from August 28 to
September 2. After data from bursts were eliminated, the rates were
divided into 400s intervals, with a pulse profile determined for each
interval. Pulse phases were then determined by correlation with the average
pulse profile. A search for sinusoidal variations was conducted for
orbital periods from $10^3$ to $10^6$ s, but no orbital modulation was
found. For periods in the range of $10^3$ s $< P_{\rm orbit} < 7.5\times
10^4$ s we can limit circular orbits to $a_{\rm x}\sin i < 0.2
(P_{\rm orbit}/10^4)^{0.15}$s (95\% confidence). For $8\times 10^3$s$ <
P_{\rm orbit} < 8\times 10^5$s, possible companion masses must be below
$0.1 M_\odot/\sin i$, assuming a 1.4$M_\odot$ neutron star.

It is likely the short-term departure of the pulse phases from the
long-term ephemeris is due to pulse shape variations. We see minor
changes in the pulse shape over the course of the August observations,
while more dramatic changes are evident between the first and second
\rxte observation. 

\subsection{Spectral Analysis}

We have performed spectral fits using XSPEC to the persistent emission
data for both \rxte observations. In both spectra we find evidence of
a line at $\sim6.7$ keV; the best fit was obtained for a powerlaw +
gaussian model. The values for the power law (photon) index, $N_H$,
and line centroid are, (2.1, 3.4$\times10^{22}$, 6.6 keV) and (3.1,
5.5$\times10^{22}$, 6.5 keV)  for June 2 and August 28, 1998,
respectively; the line is more significant in the first observation.
The power law index and $N_H$ values for June 2 are consistent with
the values reported from the \asca data (\cite{khibid}); the August 28
data exhibit a higher $N_H$ and a softer spectrum. We cannot
confidently attribute the line to \sgr as the \rxte field of view is
$\sim10$ times larger than that of \asca and may include contamination
from other sources, -- such as XTE~J$1906+09$ which was active at
least after August 30, 1998 (\cite{takeshima7008}) --, as well as from
the diffuse galactic emission.

We have further estimated the unabsorbed persistent flux between 2-10
keV to directly compare it with the \asca flux of $1.28\times10^{-11}$
ergs/cm$^2$ s reported for April 30, 1998 (\cite{khibid}). For the
June 2 and August 28, 1998 observations, the fluxes are
$\sim4.4\times10^{-11}$ ergs/cm$^2$ s and $\sim9.9\times10^{-11}$ ergs/cm$^2$
s, respectively (after taking into account the effect of the
0.441$^{\circ}$ offset of the \rxte pointing). Assuming a distance to
the source of $\sim7$ kpc (\cite{hurley94,vasisht94}), we find that
the luminosity of \sgr varied from $2.4\times10^{35}$ ergs/s to
$2.1\times10^{36}$ ergs/s, i.e., it was a factor of $\sim10$ higher
during its second active episode. Although we cannot exclude the
contribution of the diffuse galactic emission and source contamination
in the absolute values of these estimates, we believe that the ratio
of the two values yields the true luminosity increase between the two
episodes.

\section{Discussion}

We have confirmed the second case of a pulsar (\cite{khibid}) 
associated with the persistent emission of an SGR and have been able
to measure  a secular increase in its period. The first case,
SGR~$1806-20$, had a  very similar period and period derivative of
7.47 s and 8.3$\times10^{-11}$ s/s, respectively (\cite{cknature}). We
have argued that the secular spindown of SGR~$1806-20$ was due to
torques from magnetic dipole radiation and a relativistic wind, and we
have estimated the pulsar characteristic spindown age and (dipolar)
magnetic field  strength to be $\sim 1500$ years and $8\times10^{14}$
G, respectively. Our earlier observations demonstrated the existence
of `magnetars' -- neutron stars with superstrong magnetic fields
(\cite{duncan92}) -- and supported models attributing SGR bursts to
crustquakes produced  by magnetic stress (\cite{thompson95}).

In view of the strong similarities between SGRs, it would appear that
in \sgr the spindown is also the result of a magnetized wind. The
almost constant   $\dot P \simeq 1.1\times 10^{-10}$ observed in \sgr
between the May and August 1998 \rxte observations (see Figure 3)
implies  $B_\star = 8\times 10^{14}$ G, if the vacuum magnetic torque
were the only  torque acting on the star.  The age of \sgr would be
$P/2\dot P \sim 700$ years, i.e., so short that one must give up any
physical association between it and the nearby SNR G~$42.8+0.6$. 
First, the remnant is much older and, second, a tangential velocity of
$5\times 10^4\,D_7$ km s$^{-1}$ would be required at a distance of
$7\, D_7$ kpc.   The smaller $\dot P =  5.5\times 10^{-11}$ s/s
between the \asca and the \rxte May 1998 observations could be
ascribed to a glitch of magnitude $\Delta P/P = -4\times 10^{-5}$
during that interval.  Large glitches in the range $|\Delta P/P| =
10^{-5} - 10^{-4}$ are indeed implied by an extrapolation of the
glitching behavior of young pulsars to the much slower spins of the
SGR and the `anomalous X-ray pulsar' (AXP) sources
(\cite{thompson96}); but it is surpising that \sgr should not also
have been observed to glitch during the great burst of activity
associated with the August 27 event.
    
The spindown rate of a magnetar may, however, be grossly
underestimated by the vacuum dipole formalism.  The dipole luminosity
$L_{MDR} = 1.0\times 10^{34} (B_\star/10 B_{QED})^2 (R_\star/10~{\rm
km})^6  (P/{\rm 5.16~s})^{-4}$ erg s$^{-1}$ is much smaller than the
quiescent X-ray luminosity  ($L_X \sim 10^{35}-10^{36}$ erg s$^{-1}$
for both the AXP and SGR sources)  as well as the time-averaged burst
luminosity ($\langle L_{X}\rangle \sim 3\times 10^{35}$ erg s$^{-1}$
during an active phase of SGR~$1806-20$) (\cite{cknature}). In this
context, quiescent particle emission driven by internal seismic
activity increases the spindown luminosity to $I\omega \dot\omega =
L_{MDR}\times (L_{\rm particle}/L_{MDR})^{1/2}$ (\cite{thompson98})
and the simple estimate of the magnetic field derived above from $P$
and $\dot P$ is an overestimate. This effect is especially important
for  SGR~$1806-20$, where $L_{\rm particle} \sim 10^{37}$ erg s$^{-1}$
is inferred from observations of the surrounding radio halo
(\cite{vasisht95}) on both small and large angular scales
(\cite{thompson96});  for that source the required value of $B_\star$
is reduced to $2\times 10^{14}$ G. 

Transport of a magnetic field through the core and rigid crust of a
neutron star leads to two basic modes of energy release
(\cite{thompson96}): internal frictional heat which is partially
conducted to the surface;  and frequent low energy fractures that
couple to magnetospheric Alfv\'en modes and thence to energetic
particles.  In this model, an SGR burst is triggered only if the
Alfv\'en mode has a large enough amplitude to cascade to high
wavenumber before leaking out of the  magnetosphere
(\cite{thompson98}).

If \sgr turns on as a quiescent particle source ($L_{\rm particle} >
L_{MDR}$) during brief periods of burst activity, then its true age
exceeds our measured value of $P/2\dot{P}$, and a physical association
with SNR G~$42.8+0.6$  (estimated age $\sim 10^4$ yr;
\cite{vasisht95}) becomes possible. A magnetic field $B_\star = 
2\times 10^{14} t_4^{-1/2}$ G  is required for dipole radiation alone
to spin down the star to a 5.16 s period at an age of $t_4\times 10^4$
yr.   Given this value of $B_\star$, the period derivative is raised
above its long term average to $\dot P = 1.1\times 10^{-10}$ s/s if
$L_{particle} = 1.7\times 10^{36}\,(\dot P/10^{-10})^2 t_4$ erg
s$^{-1}$ -- well below the particle luminosity inferred for
SGR~$1806-20$. The total energy injected in relativistic particles
between the two \rxte observations is $\sim 2\times 10^{43}$ erg (not
including any particles injected during the August 27 event).  
Comparison with the energy of the 1$^{\prime\prime}$ synchrotron
bubble around SGR~$1806-20$ (\cite{vasisht95}) suggests that radio
flux may be detectable (at the 0.1 mJy level).   At this more advanced
age,  the tangential velocity needed to propel \sgr out of SNR
G~$42.8+0.6$ is $3400 \, D_7 \, t_4^{-1}$ km s$^{-1}$ \. Several
mechanisms could impart unusually high velocities to magnetars at
birth (\cite{duncan92}).  

The time dependence of $\dot P$ and quiescent $L_X$ in \sgr is
mirrored in the behavior of the AXP sources.  As is the case for
SGR~$1806-20$, the value of $P/\dot{P}$ for \sgr is smaller than any
of the (five) values measured so far\footnote{The spindown behavior of
1E~$2259+586$, the best-studied AXP, is discussed most recently in
\cite{baykal98}.} for `anomalous X-ray Pulsars' (AXPs), which have also
been suggested to be magnetars (\cite{duncan95,thompson96}). This
supports the suggestion (\cite{cknature}) that AXPs are a later, less
active, phase in the evolution of SGRs.  

\acknowledgments  We acknowledge support from the following grants: 
NASA grants NAG5-2755, NAG5-3674 and NAG5-4878 (J.v.P.); NASA grants
NAG5-32490 and NAG5-4799 (C.K.); NASA grant NAG5-3100 (C.T.); Texas
Advanced Research Project grant ARP-028 (R.D.); NASA grant NAG5-4238
(M.H.F.); NASA grant NAG5-4882 (K.H.)

\clearpage


\clearpage

Figure captions

\figcaption{a. The Power Density Spectrum of the May 1998 \rxte
observations of \sgr. The highest peak in the spectrum corresponds to
the fundamental period of 5.159142 s; the three less intense peaks are
the harmonics identified in the text. b. The PDS of the August 28,
1998 \rxte observation. The two highest peaks are the fundamental
period at 5.160199 s, and its second harmonic.}

\figcaption{The epoch folded pulse profile of \sgr (2-20 keV) for a)
the May 1998 \rxte observations and b) the August 28, 1998 \rxte
observation. The plot is exhibiting two phase cycles.}

\figcaption{The evolution of $\dot{P}$ versus time since the first
period measurement of \sgr with \asca (\cite{khibid}). The time is
given in Modified Julian Days (MJDs).}

\clearpage

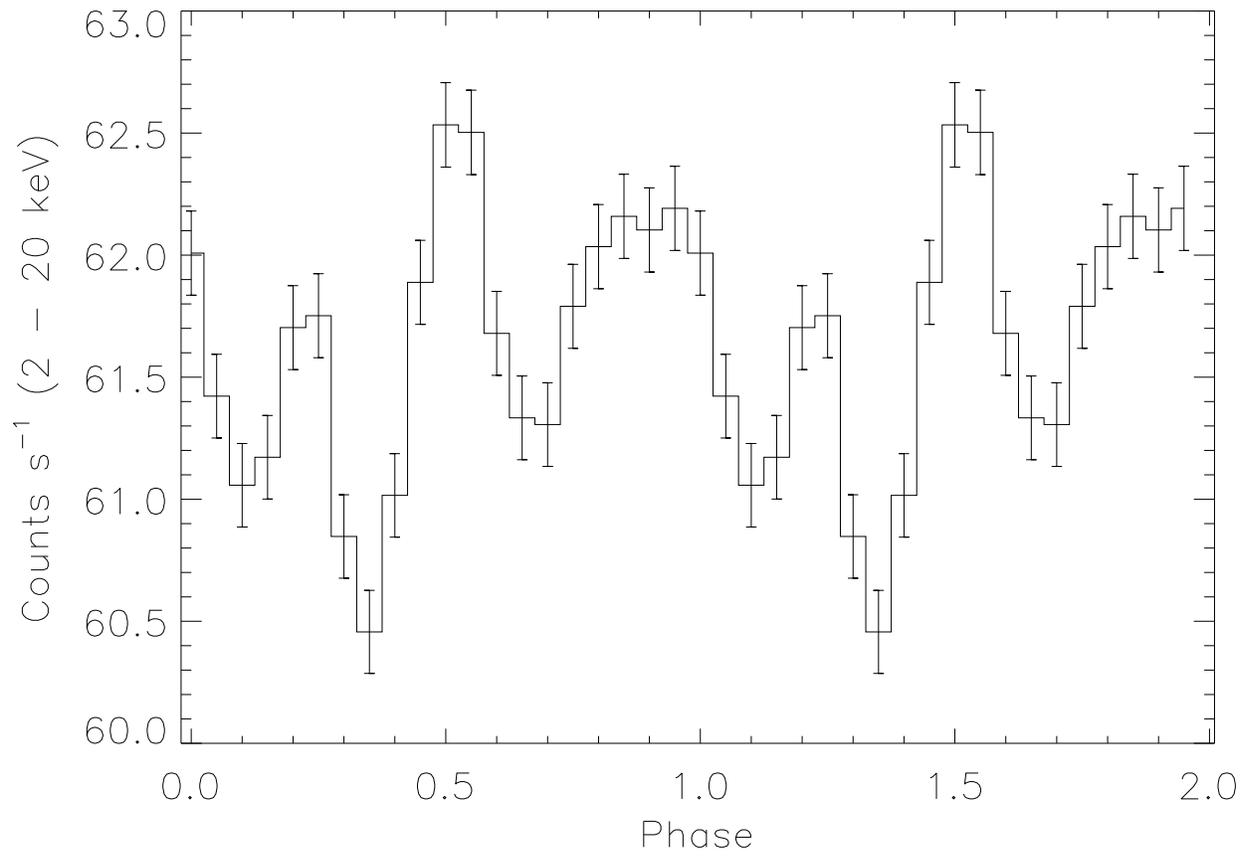
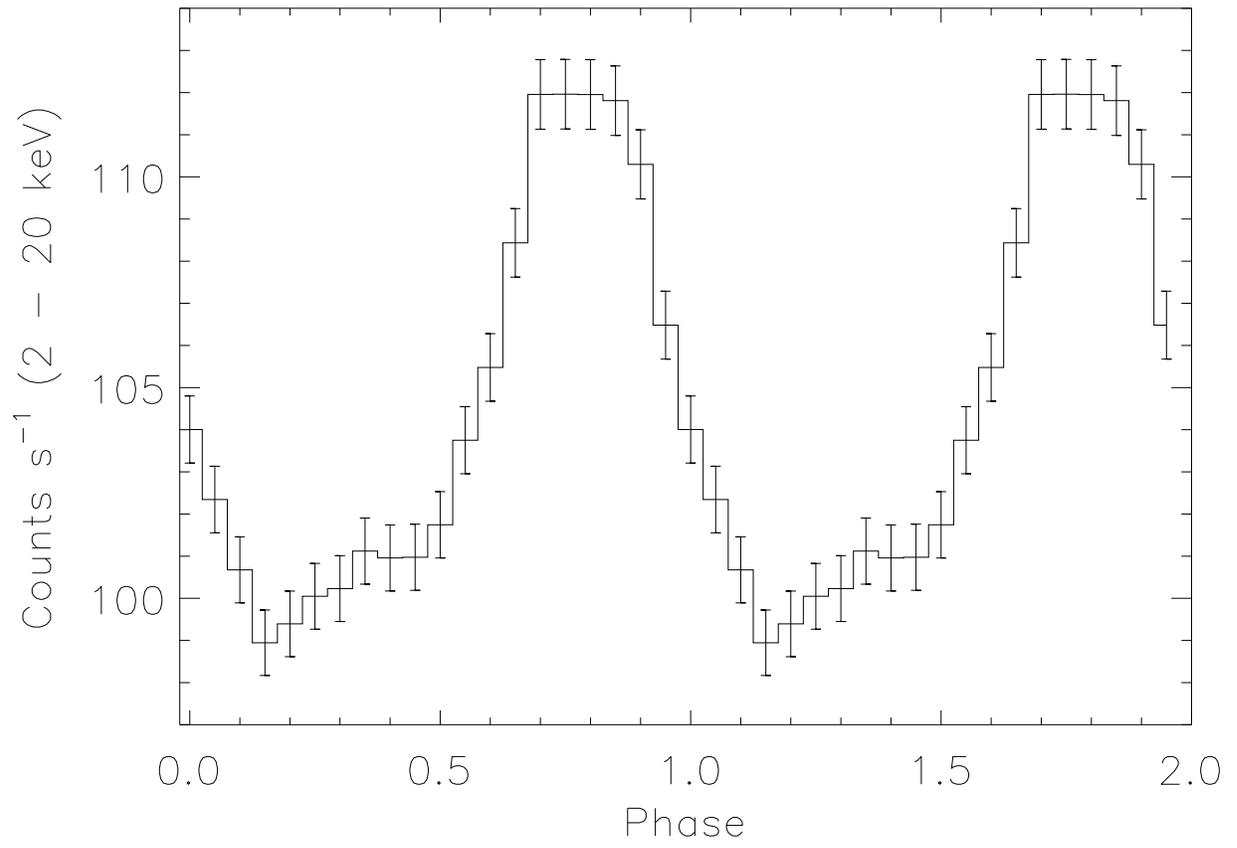
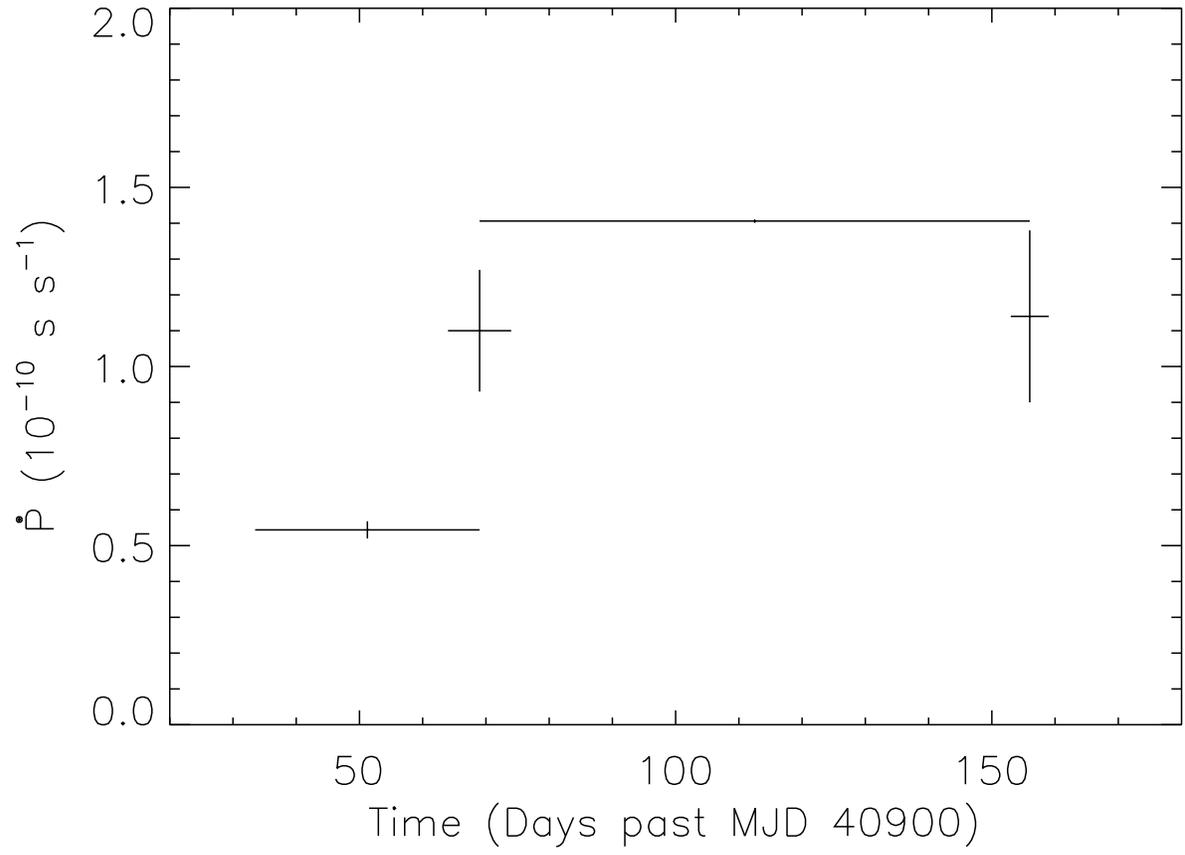
\end{document}